\documentclass[aps,prb,showpacs,twocolumn,superscriptaddress]{revtex4-2}
\usepackage{ae}
\usepackage[T1]{fontenc}
\usepackage[ansinew]{inputenc}
\usepackage{amsmath,amssymb}
\usepackage{braket}
\usepackage{subcaption}
\usepackage{appendix}
\usepackage{float}
\usepackage{color}
\usepackage{graphicx}
\usepackage[font=small, labelfont=bf,
        justification=raggedright]{caption}
\usepackage[colorlinks]{hyperref}
\usepackage{epstopdf}
\usepackage{mathrsfs}
\usepackage[normalem]{ulem}

\usepackage{soul,xcolor}
\setstcolor{red}
\usepackage{cancel}

\makeatletter
\@ifundefined{textcolor}{}
{%
 \definecolor{BLACK}{gray}{0}
 \definecolor{WHITE}{gray}{1}
 \definecolor{RED}{rgb}{1,0,0}
 \definecolor{GREEN}{rgb}{0,1,0}
 \definecolor{BLUE}{rgb}{0,0,1}
 \definecolor{CYAN}{cmyk}{1,0,0,0}
 \definecolor{MAGENTA}{cmyk}{0,1,0,0}
 \definecolor{YELLOW}{cmyk}{0,0,1,0}
\definecolor{ORANGE}{rgb}{1,0,1}
}

\@ifundefined{definecolor}
 {\@ifundefined{definecolor}
 {\@ifundefined{definecolor}
 {\@ifundefined{definecolor}
 {\@ifundefined{definecolor}
 {\usepackage{color}}{}
}{}
}{}
}{}
}{}

\usepackage{epsfig}

\begin{document}

\title{A density matrix renormalization group approach to quantum point contacts}


\author{N. Aucar Boidi}
\affiliation{The Abdus Salam International Centre for Theoretical Physics, Strada Costiera 11, I-34151, Trieste, Italy}
\email{nairaucar@gmail.com}

\author{M. N. Kiselev}
\affiliation{The Abdus Salam International Centre for Theoretical Physics, Strada Costiera 11, I-34151, Trieste, Italy}



\begin{abstract}
Using the density matrix renormalization group (DMRG) combined with the correction-vector method, we investigate the competition between an harmonic potential and repulsive interactions in a one-dimensional fermionic system. The parabolic confinement induces spatial inhomogeneity, and by tuning its curvature one can continuously interpolate between a potential well--relevant for cold-atom setups--and a quantum barrier, as realized in mesoscopic systems such as quantum point contacts.
We analyze how the ground-state particle distribution evolves with the strength and sign of the confining potential and how the confinement reshapes the spectral weight of the local density of states (LDOS) at the center of the chain. In the barrier regime, a localized peak emerges in the electron part of the spectrum ($\omega >0$) as a direct consequence of the potential. In contrast, in the well configuration and for weak interactions, a localized feature persists but shifts to the hole sector ($\omega <0$). However, for stronger interactions, the LDOS no longer displays clear signatures of the external potential, indicating that correlations dominate over single-particle confinement.
\end{abstract}


\date{\today}

\maketitle
\section{Introduction}

The study of correlated one-dimensional or quasi-one-dimensional systems has played a central role in the understanding of quantum phases of matter. Considerable effort has been devoted to characterizing low-dimensional electronic systems due to their relevance in describing physically rich scenarios, ranging from simplified toy models that capture essential mechanisms to effective descriptions of real materials \cite{giamarchi, MBcoldatoms, zoo, schubbard, wharan, PRXQSMBL,QHQPC}. The interest in these systems persists because they continuously provide new perspectives on correlation effects and collective behavior.

With the improvement of experimental techniques, the manipulation and preparation of cold-atom systems has become a realistic platform not only for testing theoretical predictions but also for engineering novel physical situations \cite{bingyang,boll,husmann,dis1,dis2}. 
In such set-ups the presence of an external confinement naturally breaks translational symmetry, rendering the system spatially inhomogeneous \cite{volosniev,matsuki}. As a consequence, the local properties become position dependent and the interplay between interactions and confinement raises fundamental questions about the resulting quantum phases and excitations \cite{greif2016,aucarcoex} as well as their transport properties for example, in Luttinger liquids~\cite{maslov,ponomarenko,safi}.

The inclusion of an harmonic potential in a one-dimensional model allows one to investigate the competition between kinetic energy, interactions, and spatial confinement. This is relevant not only for ultracold atomic gases but also for mesoscopic devices such as quantum point contacts (QPCs) \cite{volosniev}. In cold-atom systems, the harmonic trap can induce the coexistence of regions with different local character, such as metallic and Mott-insulating domains. In QPCs, the tuning of a gate-defined constriction leads to conductance quantization and reveals correlation-driven phenomena such as the 0.7 anomaly or zero-bias anomalies~\cite{mimp,qpcNature}. These systems continue to attract attention due to their role as tunable platforms for controlled quantum transport and the exploration of interaction effects in reduced dimensions. 

Among the recent directions of research, in the quantum Hall regime, QPCs continue to reveal interaction- and topology-driven effects, including anomalous conductance quantization and signatures of unconventional edge physics \cite{halfinteger,unusual,fingerprints}.
More broadly, QPCs are also used in ultracold atomic systems to probe transport, dissipation, and thermodynamic properties of superfluids under controlled conditions \cite{science2015,superfluid,arXivGiamarchi}.

In this context, the analysis of local observables becomes essential. In particular, the local density of states (LDOS) provides direct information about the spatial distribution of spectral weight and allows one to connect microscopic properties with experimentally accessible signatures \cite{qpcNature}.

The study of strong correlations in inhomogeneous systems represents a challenging task. While several numerical methods allow for indirect access to spectral properties, the Density Matrix Renormalization Group (DMRG) offers a reliable and accurate treatment of strongly correlated one-dimensional systems \cite{white,schollwoeck}. Moreover, when combined with the correction-vector approach, it enables the calculation of dynamical quantities directly in the real-frequency domain outperforming other methods that rely on indirect reconstructions of the spectrum \cite{ramasesha1,alvarez_nocera}.

In this work we investigate a one-dimensional fermionic system in which the central region includes local Coulomb interactions and is subject to an external harmonic potential, while the extremes of the chain represent non-interacting metallic leads. By computing density profiles and the local density of states, we analyze how confinement and strong correlations shape the spatial and spectral properties of the system.

The paper is organized as follows. Section~\ref{m&m} introduces the model and the numerical methods employed. In Sec.~\ref{results}, we present results for different strengths of the confining potential and interaction. First, we discuss the density profiles, and then we analyze dynamical correlators and local densities of states. Finally, Sec.~\ref{d&c} summarizes and discusses our conclusions.


\section{Models and method}\label{m&m}

In what follows, we present the specific model investigated in this work. We aim to understand the interplay of confining parabolic potentials, which define an inhomogeneous fermionic system, and electron-electron interactions. We study finite systems and we choose to use the minimum possible amount of parameters that allows to reproduce the main features of different physical situations. 

We consider a one-dimensional Hamiltonian with an interacting region only around the center of the chain, in the presence of a parabolic potential, this region is referred as the inhomogeneous interacting region (IIR) in the rest of the manuscript. The extremes of the chain represent non-interacting leads.
The interactions in the IIR, as well as the external potential, are local. We present the Hamiltonian in the following way

\begin{equation}
H = H_t + H_{inhom} + H_U,
\end{equation}%
where the different terms account for: The regular tight-binding chain
\begin{equation}
H_t = - t^{} \sum_{j,\sigma} \left( c^{\dagger}_{j+1,\sigma}c^{}_{j\sigma} + h.c. \right),
\end{equation}%
where the operator $c^{\dagger}_{j,\sigma}$ ($c^{}_{j\sigma}$) creates (destroys) a fermion with spin $\sigma$ at site $j$ and $t$ is the hopping between nearest neighbors;
the diagonal part of the Hamiltonian includes the global chemical potential $\mu$ and the site-dependent potential $\varepsilon_j$, 
\begin{equation}
H_{inhom} = \displaystyle \sum_{j,\sigma} \left(\varepsilon^{}_{j}-\mu\right) n^{}_{j\sigma},
\end{equation}%
being $n_{j,\sigma}= c^{\dagger}_{j,\sigma}c^{}_{j,\sigma}$ the number operator.
The parameter $\varepsilon_j$ defines the type of confinement present in the system being zero in the leads. 
The last term is the interacting part of the Hamiltonian:
\begin{equation}
H_U~=\sum_{j} U_j n^{}_{j\uparrow}n^{}_{j\downarrow},
\end{equation}%
where $U_j$ is the onsite Coulomb repulsion (density-density interaction) which, following Ref.~\cite{qpcNature}, 
takes the uniform value $U$ through the IIR but drops smoothly to zero near its edges and it is zero in the leads.
The non-interacting leads in the extremes of the systems are governed by the tight-binding term of the Hamiltonian, coupled to the IIR with the same hopping amplitude $t$.

All of the following results are obtained using the matrix product states (MPS) implementation of the DMRG method.  
The LDOS, 

\begin{equation}
A_{j}(\omega) = -\frac{1}{\pi} \mathbb{I}\mbox{m} \left[ G^>_{j}(\omega) + G^<_{j}(-\omega) \right],
\end{equation} is obtained using the correction vector method through the calculation of the greater and lesser Green's functions defined as

\begin{equation}
\begin{split}
G^>_j\left( \omega \right) =  \sum_\sigma  \braket{c^{}_{j\sigma} \left( \omega + i\eta - H + \epsilon_0 \right)^{-1}c^{\dagger}_{j\sigma}},  \\
G^<_j\left( \omega \right) = \sum_\sigma  \braket{c^{\dagger}_{j\sigma} \left( \omega + i\eta - H + \epsilon_0 \right)^{-1}c^{}_{j\sigma}}.
\end{split}
\end{equation} %
The energy of the ground state is $\epsilon_0$ and $\eta$ is the Lorentzian broadening.
For all the calculations, using open boundary conditions, we retain a maximum of $\sim$1000 states, we use $4-6$ sweeps and choose $\eta=0.15$.


\subsection{Barrier potential}

When we have a barrier-like potential, the system can be used to study the physics of the first subband of a QPC giving microscopic information about experimental phenomena like the so-called 0.7 anomaly or the zero-bias peak in the conductance \cite{qpcNature,fRGvonDelft,Liu}.
So the symmetric potential with respect to the geometrical center of the system $j_c$, reads

\begin{equation}
\varepsilon_j = \frac{c}{\cosh^2\left(\frac{ j-j_c}{L^\prime}\right)} \hspace{1cm} \forall j \in \text{IIR,}
\label{potqpc}
\end{equation}%
where we choose $L^\prime=(L_{IIR}-1)/4$. 
The maximum in the center of the system is defined by $c$ which, together with the choice of $L^\prime$, controls also the curvature of the potential. 
Keeping $c$ fixed and increasing $L^\prime$ will increase the curvature and thus, the potential will be smoother. The choice of Eq.~(\ref{potqpc}) was made in order to have a potential that goes smoothly to zero when reaching the extremes of the IIR and its Taylor expansion in the vicinity of the center results in
$\varepsilon_j \approx c - \frac{c}{L^{\prime 2}} \left( j-j_c \right)^2$,
so the potential is parabolic near the top. To observe conductance quantization, the regime typically considered, the potential must be sufficiently smooth \cite{glazman88} on the scale of the lattice energy $t$, so that reflection is exponentially suppressed and backscattering is negligible. The parameter $c$ is therefore chosen accordingly \cite{comCurv}. 


\subsection{Quantum well}
For the quantum well, with a minimum in the center of the chain, we choose a purely parabolic potential of the form

\begin{equation}
\varepsilon_j = -V_c + \frac{V_c}{L_*^2} (j-j_c)^2 \hspace{1cm} \forall j \in \text{IIR.}
\label{eqparabola}
\end{equation}%
Here, $V_c$ controls the height of the maximum and the curvature along with the length of the IIR. In this case when $|j-j_c|$ takes the value $L_*=(L_{IIR}+1)/2$ the potential is zero, i.e. in the boundary with the leads; then it remains zero for the rest of the sites representing the leads. To change the curvature without changing the height of the potential we can keep $V_c$ fixed and change $L_*$ (same as for the barrier). For any $V_c>0$ this potential takes negative values in the IIR.


\section{Results}\label{results}
In this section we present results for the two cases considered: The smooth barrier and the parabolic well. The first section shows static calculations: Density profiles for the ground state of the system. The second section shows the frequency dependent LDOS for one of the central sites of the chain. In both subsections we analyze the effect of changing $c$ and $V_c$ respectively for two different values of the interaction $U$. In the last part, we show how the LDOS changes along the interacting region for particular choices of the potential and both weak and strong interactions.

\subsection{Density profiles}
\begin{figure}[h!]
\centering
\includegraphics[width=0.4\textwidth]{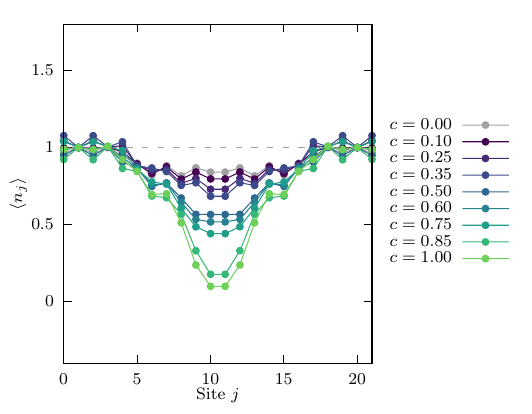}
\caption{Local occupation for the ground state in the case $L=22$, $L_{IIR}=14$, $t=U=0.5$ for different choices of $c$.}
\label{njqpcU0.5}
\end{figure}

\begin{figure}[h!]
\centering
\includegraphics[width=0.4\textwidth]{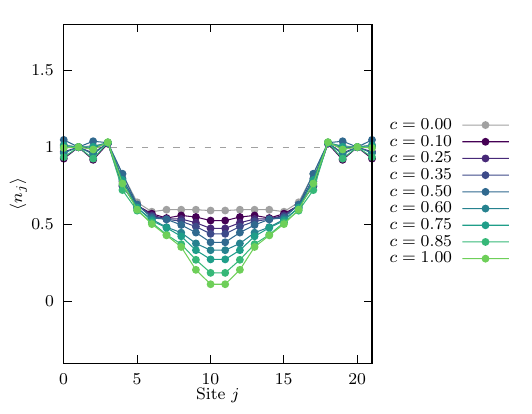}
\caption{Local occupation for the ground state in the case $L=22$, $L_{IIR}=14$, $t=0.5$, $U=4.0$ for different choices of $c$.}
\label{njqpcU4.0}
\end{figure}

\textit{Smooth barrier:}
Considering the potential in Eq.~(\ref{potqpc}), the increase of the site energy in the central region leads naturally to a lower probability of finding particles there. 
This is shown in Figs.~\ref{njqpcU0.5} and \ref{njqpcU4.0}. For both cases shown below the size of the leads is small ($L_{lead}=4$) because there are no qualitatively differences in the LDOS (shown in the following subsection) in the center of the chain when we use larger leads. 


Comparing the interacting region in both cases ($U=0.5$ and $U=4$) it is clear that a larger interaction leads to a more uniform density distributions around the center of the chain and thus, to a flatten profile, consistent with Ref.~\cite{fRGvonDelft}. When $c$ tends to be small, a larger $U$ implies lower occupation throughout the IIR than for a smaller $U$. When $c$ increases, the difference in occupation around the center between $U=0.5$ and $U=4.0$ becomes smaller but nevertheless the larger the interaction the more uniform and smooth the distribution of particles is.

The effective interaction $U_j^\text{eff} \sim U_j \cdot A_j^0(\mu)$~\cite{fRGvonDelft}, where $A_j^0(\mu)$ is the non-interacting LDOS at the Fermi level,
explains the differences between the profiles for $U=0.5$ and $U=4.0$. 
For sites close to the leads 
$U_j$ is very small in both cases, so $U^\text{eff}_j$ should also be small. The main differences appear around the center of the IIR. If the interaction $U$ is strong enough it dominates over the change of the LDOS in each site due to the potential. So in the latter, the effective interaction does not get modified by the potential. When $U$ is rather small, then the change in $A^0_j(\mu)$ over the sites will be more important, specially when $c$ is large, as in the $U=0.5$ profiles. Is also important to note that, when $c$ is large, the occupation at the center is very low, this implies a large $A^0_j(\mu)$ due to the chemical potential close to a maximum corresponding to a van Hove singularity, so here we expect similar effective interactions. 

\textit{Parabolic well:}
For a purely parabolic potential, that decreases towards the center of the interacting region, we expect the inverse behavior with respect to the previous section.
The total length of the system for the following results is $L=34$, the extension of the IIR is still 14 sites, $t=0.5$ and $\mu=0$. In all the cases, the parameter $V_c\leq 2$.

\begin{figure}[h!]
\centering
\includegraphics[width=0.4\textwidth]{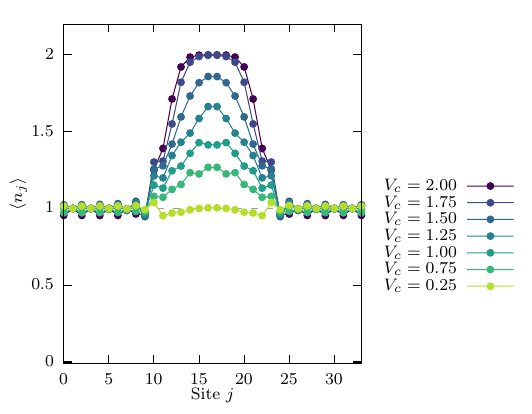}
\caption{Local occupation for the ground state in the case $L=34$, $L_{IIR}=14$, $t=U=0.5$ for different choices of $V_c$.}
\label{njparabolicU0.5}
\end{figure}

Because of the presence of this potential, the probability of finding particles in the central region of the system increases and we find a competition between the different terms in the Hamiltonian. If the electron-electron interaction is rather weak ($U=0.5$, Fig.~\ref{njparabolicU0.5}) and the site-energy due to the confining potential is large enough, we find mainly doubly occupied sites around $j_c$ as seen for $V_c=2$ and $V_c=1.75$ and then a decrease of the occupation profile towards the extremes of the IIR. 
When $V_c$ is decreased, the effect of the potential also decreases and the occupation decreases as well until it reaches values around 1. These results are consistent with the behavior found in 2D traps in cold-atom systems~\cite{greif2016}. 

\begin{figure}[h!]
\centering
\includegraphics[width=0.4\textwidth]{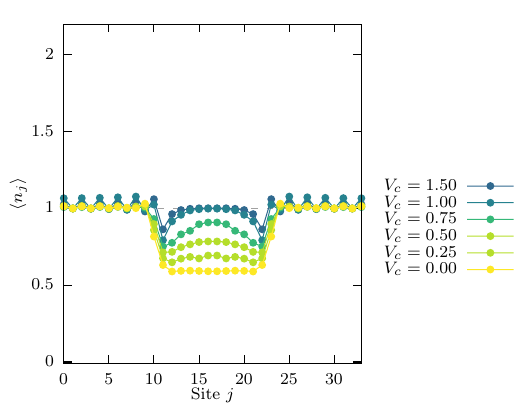}
\caption{Local occupation for the ground state in the case $L=34$, $L_{IIR}=14$, $t=0.5$, $U=4.0$ for different choices of $V_c$.}
\label{njparabolicU4.0}
\end{figure}

For stronger interactions, the results for the local occupation in case $U=4.0$ are shown in Fig.~\ref{njparabolicU4.0}. Here, even for larger values of $V_c$, the density is uniform in the sites where the potential is maximum and forms a sort of plateau around 1. Due to the strength of the interaction, here we expect to have a region with a Mott-like behavior, i.e., single occupation and antiferromagnetic nearest-neighbor correlations (not shown). In the same spirit as in Fig.~\ref{njqpcU4.0} we expect to have a greater effective interaction $U^\text{eff}$ that leads to this flatten profile for larger values of $U$.

The results of the density profiles provide an essential characterization of how an interacting system responds to the presence of a parabolic confining potential. These static observables encode different aspects of the low-energy physics of the problem, such as the charge redistribution or the emergence of spatial inhomogeneities. In particular, regions where the ground state accommodates particles most efficiently or particular microscopic configurations like regions where we can find Mott-like behavior. As a consequence, the density profiles offer clear guidance for the interpretation of the LDOS presented below.

\subsection{Density of states}
In the following subsections, we present the results of the LDOS in the central site of the IIR of the system for $U=0.5$ and $U=4.0$ interactions in both the barrier and the quantum well situations. 
 These analysis provide direct access to the local spectral reconstruction induced by the potential and interactions, and how is the competition between them. While this quantity does not directly give transport information for inhomogeneous systems, it captures local resonances that will control, for example, the low-energy conductance. 

\begin{figure}[h!]
\centering
\includegraphics[width=0.48\textwidth]{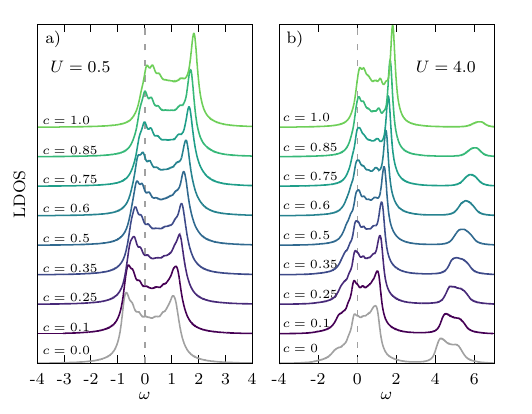}
\caption{Results for the LDOS calculated in the center of the inhomogeneous region displaced vertically, using total length $L=22$, $L_{IIR}=14$, $t=0.5$, $\mu=0$ for a) $U=0.5$ and b) $U=4.0$ and different choices of $c$.}
\label{ldosnew}
\end{figure}

\begin{figure}[h!]
\hspace*{-0.6cm}
\includegraphics[width=0.48\textwidth]{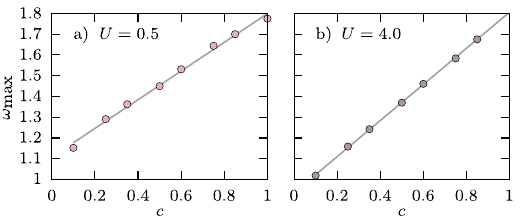}
\caption{Energy of the maximum in the LDOS, $\omega_\text{max}$, as a function of $c$ for a) $U=0.5$ and b) $U=4.0$. Solid grey lines are the linear fittings for each case.}
\label{wmax}
\end{figure}

\textit{Smooth barrier:}
We calculate the LDOS for $L=22$, $L_{IIR}=14$, $t=0.5$ and $\mu=0$. Figure~\ref{ldosnew}a shows the case of $U=0.5$, an interaction relatively small, for different choices of $c$ starting from the bottom with the reference case of $c=0$. Note that this case is not symmetric with respect to $\omega=0$ due to the presence of interactions in the central region and thus a slightly decrease in probability of finding particles here. For small enough values of $c$, the results show a LDOS that resembles the LDOS of a simple tight-binding model, with rather symmetric van Hove peaks in the extremes of the structure. When $c$ is increased, the whole structure moves slightly to the right, indicating a decrease in the local density, and developing an asymmetric shape.
The higher-energy peak in the boundary of the structure at $\omega_\text{max}(c)\sim 2t+\mathcal{O}(c)$ (Fig.~\ref{wmax}a), gains weight being a direct consequence of the presence of the external potential. 


For stronger interactions, Fig.~\ref{ldosnew} b) shows, from the bottom to the top, how the structure of the LDOS depends on the parameter $c$. For $c=0$, the grey curve shows the typical LDOS of a hole-doped uniform Hubbard model. The structure around $\omega=0$ is related to excitations whose configurations consist in single occupied sites, while the structure around $\omega=5$ gives information about excitations related to doublon configurations. The evolution of the LDOS reflects two things: The transference of spectral weight typical of a Hubbard model, an indication of strong correlations, and the effect of the parabolic potential. 

Again, the peak that comes from the potential 
moves towards higher energies when $c$ is increased (Fig.~\ref{wmax}b)
and the shape of the LDOS changes until its high-energy structure is almost negligible. In the latter case, $c=1$, the main structure corresponds to adding a particle to an almost empty central region, reflecting only the presence of the confining potential (similar to the case $U=0.5$, $c=1$).

\textit{Parabolic well:}
For the parabolic well, we present results for the LDOS showing how the local occupation increases. For the different strengths of the interaction $U$ we observe how the shape of the LDOS reflects the presence of the parabolic potential.

\begin{figure}[h!]
\hspace*{-0.6cm}
\includegraphics[width=0.48\textwidth]{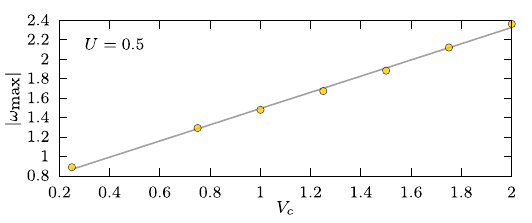}
\caption{Absolute value of the energy of the maximum in the LDOS, $|\omega_\text{max}|$, as a function of $V_c$ for $U=0.5$, the solid grey lines is the linear fitting for the data.}
\label{wmaxParab}
\end{figure}

For small $U=0.5$, the maximum in the LDOS appears for negative values of $\omega$, the shape of the LDOS is basically the same as for the potential in Eq.~(\ref{potqpc}) but reflected with respect to $\omega=0$, highlighting the idea that the precise parametrization of the potential does not play a crucial role but the shape of the potential in the top should be parabolic \cite{qpcNature}. Figure~\ref{wmaxParab} shows that the position of the maximum increases monotonically with the value of $V_c$. 

\begin{figure}[h!]
\centering
\includegraphics[width=0.45\textwidth]{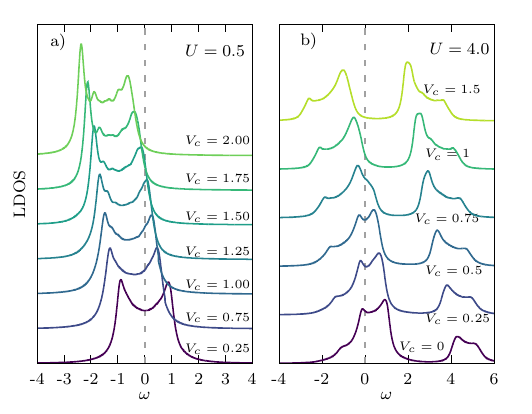}
\caption{Results for the LDOS calculated in the center of the IIR for the purely parabolic potential defined in Eq.~(\ref{eqparabola}). The parameters used are $L=34$, $L_{IIR}=14$, $t=0.5$ and $\mu=0$. The left (right) panel shows results for $U=0.5$ ($U=4.0$).}
\label{ldosparabolic}
\end{figure}

Again from the bottom to the top, for each choice of $U$ we show the two known results, case $V_c=0$. For $U=0.5$ the LDOS again resembles the tight-binding DOS, while for $U=4.0$ we find the hole-doped Hubbard result. In the latter, as we decrease $V_c$ we move to a Mott-insulating region so the shape of the two structures tends to be equal. 

Considering a homogeneous Hubbard model without explicit particle-hole symmetry 
when the chemical potential is zero the system will be around quarter filling (case $V_c=0$). Then if we change the potential between $2t$ and $U-2t$, we will be inside the Mott gap and the system will be half filled. So, by considering the local chemical potential, given in this case by the confining potential and the global chemical potential (the latter chosen equal to zero for these cases), one can predict what kind of structure will the LDOS have in that region and thus, predict its physical nature. Here again, we clearly see two structures, the left structure can be related to the LHB and the right one the UHB. 
Nevertheless, for these parameters, the effect of the parabolic potential seems to be negligible in the shape of the LDOS.
  If we continue increasing the value of $V_c$, we will have a region above half filling, the high-energy structure is going to move towards lower energies, crossing $\omega=0$. In this case, the effect of $V_c$ is going to reveal in the lower extreme of this band, in the same sense that the results for $U=4$ for the smooth barrier.

\subsection{Heatmaps: Site-dependent density of states}
\begin{figure}[h!]
\hspace*{-1.2cm}
\begin{subfigure}{0.6\textwidth}
\includegraphics[width=1\textwidth]{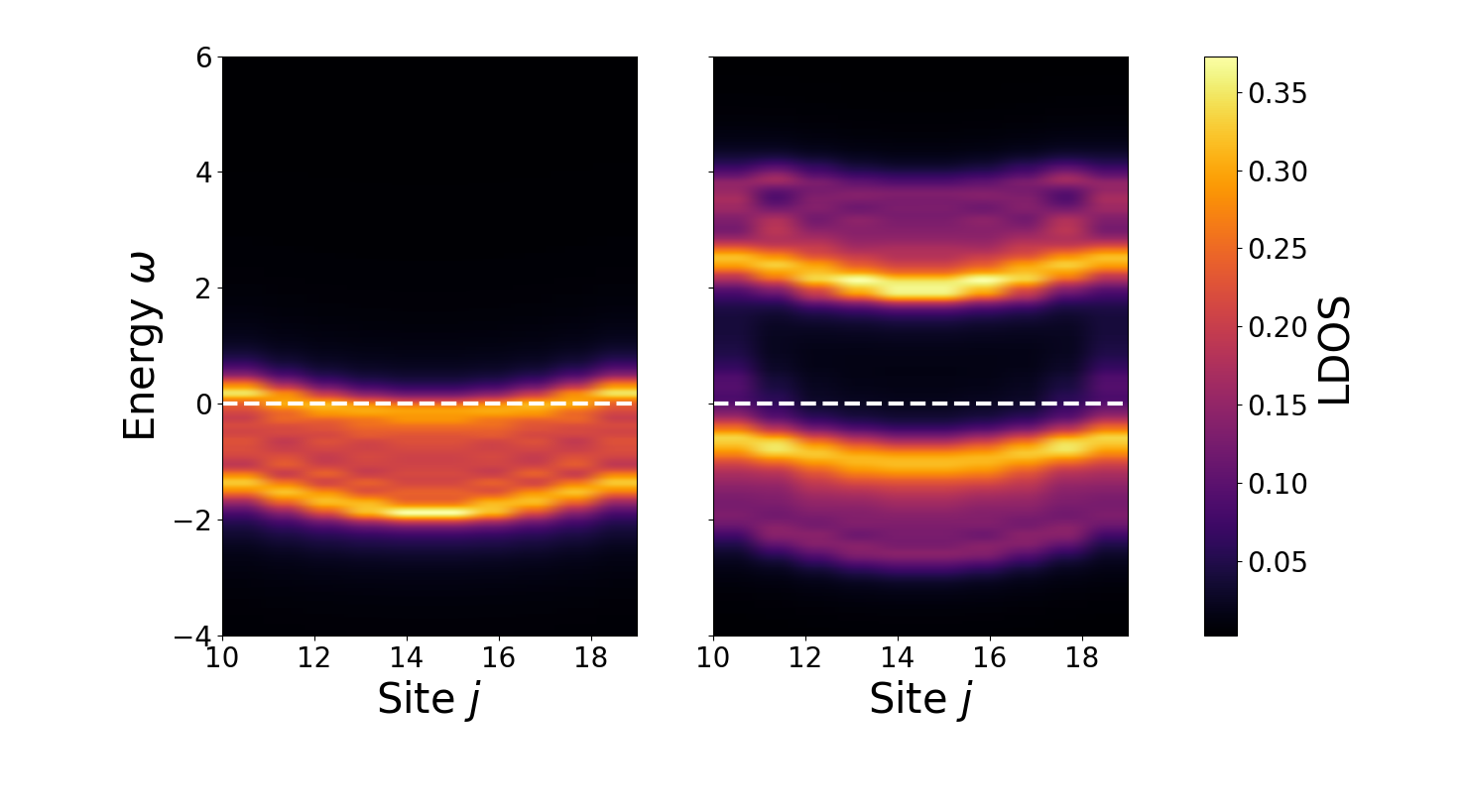}
\end{subfigure}
\hspace*{-1.2cm}\begin{subfigure}{0.6\textwidth}
\includegraphics[width=1\textwidth]{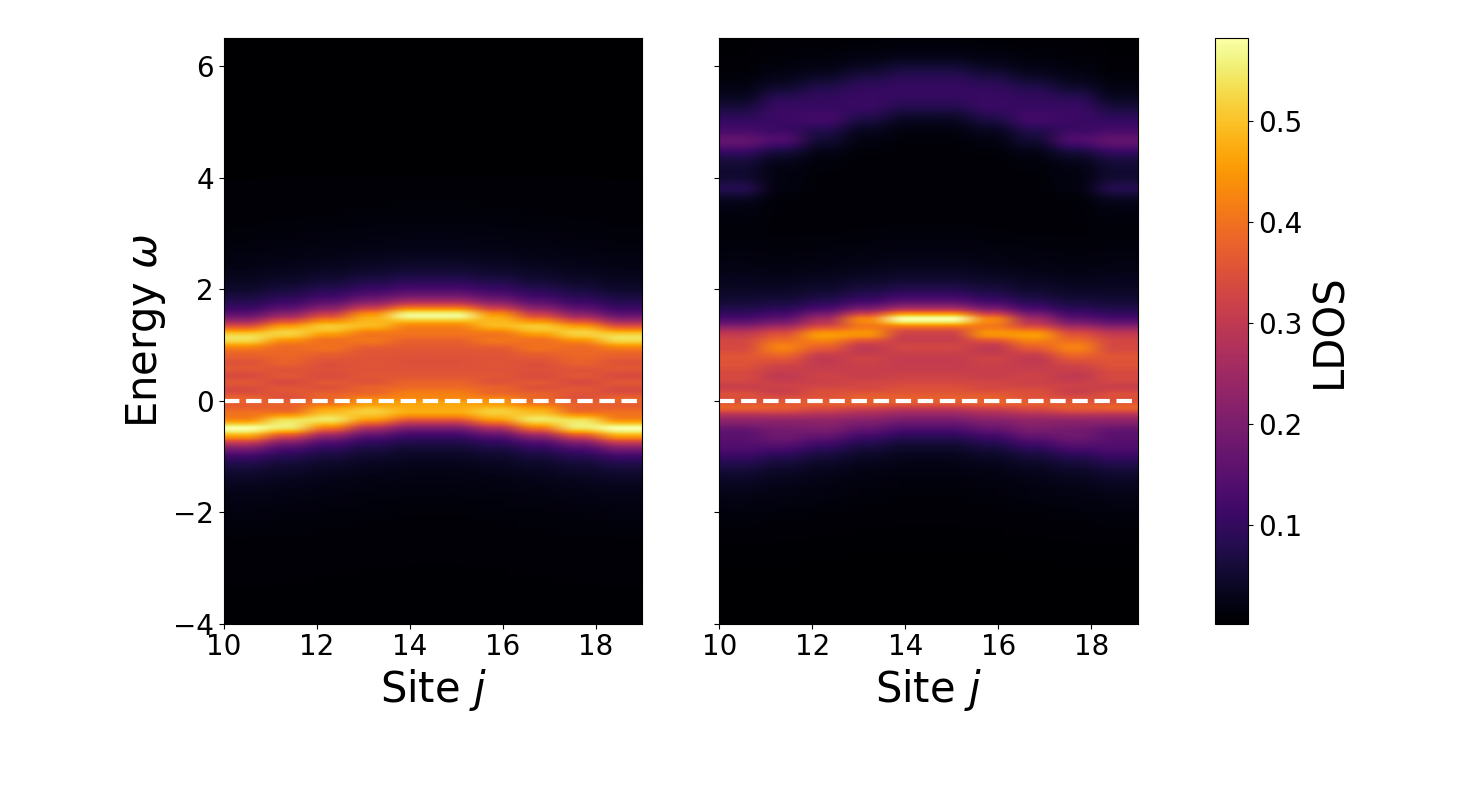}
\end{subfigure}
\caption{Heatmaps for $L=30$, $L_{IIR}=14$, $t=0.5$, $V_c=-1.5$ and $U=0.5$ (left panels) and $U=4.0$ (right panels). a) For the well and b) for the barrier.}
\label{heatmaps}
\end{figure}

In Fig.~\ref{heatmaps} we present heatmaps for some particular cases of the barrier and the parabolic well. We show the LDOS along the IIR of the chain for $L=30$, $L_{IIR}=14$, and $t=0.5$.
While the colors indicate the weight of the LDOS, $\omega$ is shown on the vertical axis and the horizontal axis indicates the site index. The Fermi level is marked with a dashed white line in all the figures.
For all the cases, we observe that the support of the LDOS follows the shape of the external potential, being concave or convex depending on the case.

For $U=0.5$, in the left panels of Fig.~\ref{heatmaps} we show results for the well (top) and the barrier (bottom). As expected, we observe only one structure in the LDOS along the IIR, with maxima at its extremes for all sites. For the well, the peak at $\omega=-2$ which corresponds to sites close to the center of the chain has a higher weight compared to the corresponding peak in the rest of the IIR. As we move towards the leads, this maximum--now with smaller weight--also shifts to lower energies, i.e. closer to $\omega=0$. 
For the barrier, there is a maximum in the center of the chain around $\omega=2$. For the sites closer to the leads the weight of the maximums in the LDOS is very similar. This is because, due to the choice of the potential, its value for those sites is very low, so the LDOS should resemble the tight-binding LDOS with symmetric peaks for the van Hove singularities.

In the case of $U=4.0$, the right panels of Fig.~\ref{heatmaps} show the heatmaps for the well and the barrier, respectively. For both cases, we observe two structures in the site-dependent LDOS. For the well, both structures have a similar weight distribution, with maxima at energies around $\omega=-1$ and $\omega=2$, which persist along the entire IIR, indicating a Mott-insulating-like behavior. In the case of the barrier, the lower-energy structure has a much higher weight. There is a pronounced peak close to $\omega=2$, with higher weight for sites near the center, similar to the case previously described for $U=0.5$.

\section{Discussion and conclusions}\label{d&c}
In this work we studied an inhomogeneous one-dimensional interacting fermionic system coupled to non-interacting leads. The inhomogeneity is introduced through an external potential acting on the central region, chosen either as a smooth barrier or a parabolic well. Using the matrix product state implementation of the density matrix renormalization group, we analyzed density profiles and local densities of states (LDOS) mainly at the center of the interacting region. 

Our results provide a unified description of how interactions, itinerancy and confinement compete in finite inhomogeneous systems. By focusing on minimal models and relatively small system sizes, we recover and connect the physics of homogeneous interacting chains, quantum point contacts and confined cold-atom systems within a single framework.

A detailed characterization of spectral properties in inhomogeneous fermionic systems is essential for understanding the behavior of quantum point contacts and related nanostructures, as well as harmonically trapped cold atoms, which constitute controlled experimental platforms to probe many-body effects~\cite{qpcNature,fRGvonDelft,fRGKarrasch,greif2016,matsuki,MBcoldatoms}.

An important advantage of our approach is the direct access to the LDOS without requiring analytic continuation or additional projections of the Hilbert space. Although restricted to finite systems, this allows for a microscopic interpretation of the spectral features and their relation to local density configurations.

For both confining potentials considered, we find results consistent with previous studies while highlighting how characteristic features emerge already in relatively small systems. In particular, the LDOS at the center of the interacting region reveals clear signatures of the competition between interactions, kinetic energy and confinement.

When the curvature of the potential is large enough the external potential is directly reflected in the LDOS as a pronounced maximum. For the barrier case, and fillings below half filling, the position and spectral weight of this peak increase with curvature for both weak and strong interactions, while the remaining spectral features closely resemble those of the corresponding homogeneous systems.

In contrast, for a parabolic well, weak interactions lead to a peak in the hole sector that shifts in energy and weight as the curvature is changed. For strong interactions, the LDOS evolves towards a Mott-like or hole-doped Hubbard spectrum, depending on the curvature, with the confining potential playing a secondary role. In this regime, strong correlations are responsible for the formation of a Mott region around the center of the interacting region, which progressively changes towards the interfaces with the leads as the curvature is increased, these results highlight the robustness of the Mott state.

We show that a useful starting point for interpreting these results is the analysis of the corresponding homogeneous systems combined with a local shift of the chemical potential~\cite{aucarcoex,matsuki}. While this picture captures the dominant trends, additional features in the LDOS depend on the presence of the confining potential.

Finally, the analysis and characterization of the LDOS constitute a natural starting point for identifying microscopic mechanisms underlying quantum transport phenomena. In particular, features such as the van Hove ridge discussed in early studies of quantum point contacts~\cite{qpcNature} illustrate how local spectral properties provide a microscopic perspective on transport phenomena.




\section{Acknowledgements}
We thank K. Hallberg, J. von Delft and N. Sobrino for the useful discussions.


\begin{thebibliography}{99}
\bibitem{comCurv} 
The latter translates in the condition $c \ll L^{\prime 2}t/4$. In all of the results we used $t=0.5$ and the shorter length for the CCR was $L_{IIR}=14$ so the condition is fulfilled for all the choices of $c$.
\end{thebibliography}
\end{document}